\documentclass[letterpaper, 10 pt, conference]{ieeeconf}  

\IEEEoverridecommandlockouts                              

\usepackage{cite}
\usepackage{amsmath,amssymb,amsfonts}
\usepackage{algorithmic}
\usepackage{graphicx}
\usepackage{textcomp}

\usepackage{graphicx}
\usepackage{subfigure}
\usepackage{epsfig} 

\usepackage{hyperref}

\usepackage{amsthm}

\newcommand{\R}{\mathbb{R}}

\newcommand{\N}{\mathbb{N}}

\newcommand{\bmat}[1]{\begin{bmatrix}#1\end{bmatrix}}
\newcommand{\norm}[1]{\left\lVert{#1}\right\rVert}

\newcommand{\ip}[2]{\left\langle #1, #2 \right\rangle}

\newcommand{\mcl}[1]{\mathcal{ #1}}
\newcommand{\mbf}[1]{\mathbf{ #1}}

\newtheorem{thm}{Theorem}

\newtheorem{lem}[thm]{Lemma}

\newtheorem{cor}[thm]{Corollary}

\title{\LARGE \bf
A New Parametric Uncertainty Approach to Systems with Time-Varying Delay and Feedback Stabilization
}

\author{Aleksandr Talitckii$^{1}$, Eduardo S. Tognetti$^{2}$, Peter Seiler$^{3}$ and Matthew M. Peet$^{1}$ 
\thanks{This work was supported in part by the National Science Foundation under Grants NSF CCF-2323532, NSF EPCN-2337751 and NSF EPCN-2337752, and by CNPq, Brazil, under Grant 302987/2025-8. (Corresponding author: Aleksandr Talitckii)}
\thanks{$^{1}$School for the Engineering of Matter, Transport and Energy, Arizona State University, Tempe, AZ, 85298, USA. {\tt\small atalitck@asu.edu} and {\tt\small mpeet@asu.edu}}%
\thanks{$^{2}$Department of Electrical Engineering, University of Bras\'{\i}lia, 70910-900, Bras\'{\i}lia, DF, Brazil{\tt\small estognetti@unb.br}}%
\thanks{$^{3}$Department of Electrical Engineering and Computer
 Science, University of Michigan, Ann Arbor, MI 48109, USA
        {\tt\small pseiler@umich.edu}}%
}

\begin{document}

\maketitle
\thispagestyle{empty}
\pagestyle{empty}

\begin{abstract}

Delay Differential Equations (DDEs) with an uncertain time-varying state delay are often modeled as a feedback interconnection of a nominal time-invariant system with a time-varying uncertainty operator. However, typical choices of uncertainty operators do not admit tight characterizations of input-output properties. In this paper, we propose a new representation of a DDE with time-varying delay system as a feedback interconnection of a nominal fixed delay DDE with a time-varying multiplicative parametric uncertainty operator, where the parameter is bounded in $L_\infty$ norms -- yielding an operator norm bound on the uncertainty. We then provide a modified uncertainty representation which uses the $L_2$ norm of the parameter to bound the operator norm of the uncertainty. Finally, we show that by representing the nominal system using Partial Integral Equations (PIEs), these characterizations of the uncertainty operator can be used for stability analysis or stabilizing feedback controller synthesis. 
The effectiveness of the approach is illustrated with simple numerical examples.  

\end{abstract}

\section{Introduction}
\label{sec:introduction}

Recent advances in Delay Differential Equations (DDEs) provide many methods for stabilizing systems with constant delays -- using e.g. Lyapunov-Krasovskii functionals~\cite{kharitonov2003lyapunov,gu2003stability}; frequency domain methods~\cite{niculescu2003delay}; multiple model transformations~\cite{fridman2002descriptor,kharitonov2003additional}; first-order hyperbolic Partial Differential Equations (PDEs)~\cite{karafyllis2014relation, krstic2008backstepping}; and, probably the most recent, Partial Integral Equations (PIEs)~\cite{peet2021representation}. However, for systems with time-varying delays, further attempts to develop these methods have been limited. A typical approach is to consider the DDE with time-varying delays as an interconnection between a system with a constant delay and an uncertainty operator~\cite{kao2007stability}. Then, by analyzing the input-output properties of these subsystems,  sufficient conditions  for the interconnected system to design a stabilizing controller have been developed in e.g.~\cite{fridman2006new}. 

To represent the time-varying delay as an uncertainty, there are two most commonly used options: either the delayed signal itself, $(\Delta x)(t) = x(t - \tau(t))$, or a difference between the delayed signal and some nominal signal, $(\Delta x)(t) = x(t- \tau(t)) - x(t - \tau_0)$, where $\tau_0$ is a nominal delay.  Then, representations of the DDE as a feedback interconnection between a nominal system and one of these uncertainty operators allow us to show stability or design a controller of the interconnected system. 
However, such input-output characterizations of the uncertainty operator are not tight. For example, for a system with uncertain constant delay, $\dot x(t) = -x(t - \tau)$ (stable for $\tau < \pi/2$), bounds on the delay operator may not prove stability. This system may be represented using uncertainty operator $w(t) = (\Delta x)(t) := x(t) - x(t - \tau)$ (with $\|\Delta\|_{H_\infty} = 2$) interconnected with nominal system, $\dot x(t) = -x(t) + w(t)$, (with $H_\infty$ norm of $1$). Hence, small gain fails to prove stability for any $\tau$. For systems with time-varying delays, the dynamic nature of the delay further complicates the characterization of the uncertainty operator. Thus, input-output stability analysis is mainly restricted to $L_2$-gain conditions for robust control~\cite{fridman2006new} or a limited set of Integral Quadratic Constraints for stability analysis~\cite{kao2007stability, pfifer2015integral}.



To reduce the conservatism introduced by the uncertainty representation, we may consider alternative uncertainty operators. For example, systems with uncertain constant delays can be described using ODE-PDE or PIE state-space representations. Such models allow the system to be represented as feedback interconnection of a nominal PIE with an infinite-dimensional static parametric uncertainty~\cite{wu2021robust, talitckii2023integral}. Then, the static parametric uncertainty admits a tighter characterization in terms of multipliers and small gain. 

For systems with time-varying delay, however, there is no similar representation as a feedback interconnection with parametric uncertainty. Our first task, therefore, is to represent the DDE with time-varying delay as an interconnection of a time-invariant nominal system with a time-varying parametric uncertainty operator -- See Fig.~\ref{fig:interconnection}. This uncertainty operator is characterized by the discontinuous function $I(s,\frac{\tau(t)}{\tau_{max}})$ where $I(x,y)$ is the indicator function and $\tau_{max}$ is the maximum delay. Then, we use properties of the $L_{\infty}$ and $L_2$ norms applied to $I(\bullet,\frac{\tau(t)}{\tau_{max}})$ to bound the uncertainty in cases where the minimum delay is zero or strictly positive, respectively. 

%
%
Finally, to achieve robust stabilization for the interconnected system, we combine $H_\infty$ optimal control of the nominal PIE system with the small gain conditions as discussed in Sec.~\ref{sec:stabilization}. Next, using the equivalence of the input-output maps for the DDE and the PIE system (Thm.~\ref{thm:DDE-and-ODE-PDE-are-equivalent}), we show that if the controller stabilizes the PIE system, then  the closed-loop DDE is also stable   (Cor.~\ref{cor:stabilization1}). Furthermore, the PIE representation of the DDE allows us to consider a more general type of controllers, including the integral of the solution history for the DDE.

Finally, in Section~\ref{sec:examples} we test the proposed method for several examples and show the superiority of the proposed approach over alternative methods for stabilization of DDEs with time-varying delays. Furthermore, we discuss the choice of the uncertainty operator for the analysis of the DDE.\vspace{-4mm}

\subsection*{Notation}
We denote by $\N, \R$ and $\R_+$  the natural numbers, the real numbers and the non-negative real numbers, respectively. For $\Omega \subseteq \R$, we use $L_2^n(\Omega)$ to denote the space of Lebesgue square integrable functions $f:\Omega\rightarrow \R^n$ with inner product $\ip{f}{g} = \int_{\Omega} f(s)^T g(s)ds$. For compact $\Omega \subseteq \R$ we use $C^k(\Omega)$ to denote the space of $k$ times continuously differentiable functions $f:\Omega\rightarrow \R^n$ with norm $\|f\| = \sup_{t \in \Omega} \|f(t)\|_2$.   
For given $a, b \in \R$ and $n_1, n_2 \in \N$ we use $\mbf Z^{n_1, n_2}$ to denote the Hilbert Space $\mbf Z^{n_1, n_2} := \R^{n_1} \times L_2^{n_2}[a, b]$.  We denote by $\mcl I$ and $0$ the identity and null operator on the $\mbf Z$, respectively.



\section{Problem Formulation}
\label{subsec:problem_formulation}
In this paper,  we consider a linear DDE of the form \vspace{-2mm}
\begin{align}
\dot x(t) & = A x(t) + A_d x(t - \tau(t)) + B_u u(t) + B_w w(t),  \;  t \geq 0 \notag\\ 
e(t) & = C x(t) \label{eqn:DDE} \\ 
x(s) & = x_0(s), \qquad s\in[-\tau_{max},0 ], \qquad \tau_{max} \geq 0\notag \\[-7mm] \notag
\end{align}
where $A, A_d \in \R^{n_x\times n_x}$, $B_u \in \R^{n_x \times n_u}$, $B_w \in \R^{n_x \times n_w}$ and $C \in \R^{n_e \times n_x}$. Moreover, $\tau$ is continuous and $\tau(t) \in [\tau_{\min}, \tau_{\max}]$, where $\tau_{max} >  \tau_{min} \geq 0$ are the minimum and maximum delay margins. Finally, $x(t) \in \R^{n_x}$, $u(t) \in \R^{n_u}$, $w(t) \in \R^{n_w}$ and $e(t) \in \R^{n_e}$ are the state, control input, disturbance, and regulated output at time $t$, respectively. The initial condition is given by $x_0 \in C^1([-\tau_{max}, 0])$.

\textbf{Remark 1.} In this paper, we generally avoid stating explicit conditions for the well-posedness of the time-varying DDE. Instead, the theorems are stated so that if a suitably regular solution to the DDE exists, then this solution also satisfies a given ODE-PDE and, ultimately, a PIE. This approach allows for application to any suitably well-posed DDE. That said, sufficient conditions for well-posedness of the DDE in~\eqref{eqn:DDE} can be found in Thm 5.2 of~\cite{kolmanovskii2013introduction} as $u, w, \tau \in C(\R)$. For systems with input delay, we would require $u, w \in C^1(\R)$. Alternative conditions for well-posedness also impose the condition $\dot \tau(t) \le 1$~\cite{verriest2010well}.




The feedback controllers developed in this paper will be taken to have the form \vspace{-3mm}
\begin{equation}\label{eqn:DDE_control}
u(t) = K_1x(t) + \int_{-\tau_{max}}^0 K_2(s) \dot x(t + s) ds,\vspace{-2mm}
\end{equation}
where $K_1 \in \R^{n_u \times n_x}$ and $K_2$ is a bounded function. Note that $\dot x$ is used in the feedback to obtain a more compact parametrization of the controller. Then, any controller in the more traditional form
\vspace{-3mm}
\begin{align*}
u(t) &= K_P x(t) + K_D x(t-\tau_{max})+ \int_{-\tau_{max}}^0 \hspace{-4mm}
K_I(s) x(t+s) ds, \\[-6.5mm] 
\end{align*}
can be represented as in Eqn.~\eqref{eqn:DDE_control} with $K_2(s) = - K_D - \int_{-\tau_{max}}^s K_I(\theta)d\theta$ and $K_1=K_P - K_2(0)$. Conversely, any controller of the form in Eq.~\eqref{eqn:DDE_control} may also be represented in the traditional form using $K_P=K_1 + K_2(0)$, $K_D=-K_2(-\tau_{max})$, and $K_I(s)=-\partial_s K_2(s)$.

To design a controller as defined in Eqn.~\eqref{eqn:DDE_control}, we first consider an ODE-PDE representation for systems with time-varying delays. Then, considering the time-delay as an uncertainty, we reformulate the problem as a feedback interconnection. Finally, based on properties of the interconnected systems, we propose sufficient conditions for the stabilization of the DDE in~\eqref{eqn:DDE}.


\section{An ODE-PDE State-Space Representation for Systems with Time-Varying Delays}\label{sec:reformulation_ODEPDE}
 
 In this section, we propose an ODE-PDE representation of
 the DDE in Eqn.~\eqref{eqn:DDE}. We  accomplish this task in two steps. First, we review the ODE-PDE model for a system with a constant delay (Subsec.~\ref{subsec:ODEPDE_for_constant_delays}). Then, we represent a system with a time-varying delay using a generalized ODE-PDE model (Eqn.~\eqref{eqn:ODE-PDE-TVD} in Subsec.~\ref{subsec:ODEPDE_for_timevar_delays}), which has equivalent input-output and internal stability properties to the DDE in~\eqref{eqn:DDE} (Thm.~\ref{thm:DDE-and-ODE-PDE-are-equivalent}). \vspace{-1.5mm}

\subsection{Representation of Systems with Constant Delays}
\label{subsec:ODEPDE_for_constant_delays}
 We start with System~\eqref{eqn:DDE} assuming a constant time delay $\tau(t) = \tau_{max}$. As shown in~\cite{peet2021representation}, systems with constant delays admit an ODE-PDE state-space  representation with equivalent internal and input-output stability properties. Specifically, let $\tau(t) = \tau_{max}$ for $t \geq 0$, then the DDE in~\eqref{eqn:DDE} can be represented as follows \vspace{-4mm}
\begin{align}
 &\dot x(t)\hspace{-0.5mm} = \hspace{-0.5mm}(A\hspace{-0.5mm} +\hspace{-0.5mm} A_d) x(t) - A_d \hspace{-1mm} \int_{-1}^0 \hspace{-1mm}\partial_s \phi(t, s) ds  + B_u u(t) + B_w w(t)   \notag\\[-2mm]
& \partial_t \phi(t, s) = \frac{1}{\tau_{max}} \partial_s \phi(t,s), \quad \phi(t, 0) =  x(t) \label{eqn:ODE-PDE_constant_delay}\\
& e(t) = Cx(t), \quad  x(0) = x_0(0),  \quad \phi(0, s) = x_0(\tau_{max} s),  \notag \\[-7mm] \notag
\end{align}
where  $t \geq 0$ and $s \in [-1, 0]$. 

Then, the exact solution of the transport PDE in Eqn.~\eqref{eqn:ODE-PDE_constant_delay} is defined as  \vspace{-4mm}
\begin{equation} \label{eqn:PDE-solution}
\phi(t, s) = \begin{cases}
x(t + \tau_{max} s), & \text{ if } t + \tau_{max} s \geq 0\\
x_0(t + \tau_{max} s), & \text{ otherwise }
\end{cases}
\end{equation}
for all $t \geq 0$ and $s \in [-1, 0]$. 

Furthermore, if there exists a solution of the ODE-PDE system, we are able to reconstruct the solution of DDE~\eqref{eqn:DDE} with $\tau(t) = \tau_{max}$ as \vspace{-4mm}
\begin{equation} \label{eqn:DDE-solution-from-ODEPDE}
\hat x(t) = \begin{cases}
x(t), & \text{ if } t \geq 0\\
x_0(t), & \text{ otherwise. }
\end{cases}\vspace{-2mm}
\end{equation}
Then, the equivalence between DDE~\eqref{eqn:DDE} with constant time-delay and ODE-PDE~\eqref{eqn:ODE-PDE_constant_delay} was established as follows.\vspace{-1.5mm}
\begin{thm}[Lemmas~1 and~3 in~\cite{peet2021representation}]\label{thm:ODE-PDE-and-DDE-with-constant-delay}
Suppose $\tau(t) = \tau_{max}$ for all $t \geq 0$, $u, w \in C(\R_+)$ and $x_0 \in C^1([-\tau_{max}, 0])$.   Then, $x \in C^1([-\tau_{max}, \infty))$ satisfies DDE~\eqref{eqn:DDE} with inputs $u, w$ and initial conditions $x_0$ if and only if $x \in C^1(\R_+)$ and $\phi \in C^1(\R_+ \times [-1, 0])$ satisfy ODE-PDE~\eqref{eqn:ODE-PDE_constant_delay} with inputs $u, w$ and initial conditions as in Eqn.~\eqref{eqn:ODE-PDE_constant_delay}. 
\end{thm}\vspace{-1.5mm}




Note that a classical result of the equivalence between the DDE and the ODE-PDE systems requires differentiability $x$ at $t = 0$ (the ``sewing condition'')~\cite{peet2021representation}. The assumption $x \in C^1([-\tau_{max}, 0))$ in Thm.~\ref{thm:ODE-PDE-and-DDE-with-constant-delay} ensures that the ``sewing condition'' is satisfied.





Although the ODE-PDE representation has been successfully applied to systems with uncertain but constant delays~\cite{zhu2015adaptive, wu2021robust}, the ODE-PDE representation has not been extended to time-varying delays. In the next subsection, we present an ODE-PDE model with time-varying parameters that allows us to equivalently represent systems with time-varying delays.

\subsection{Representation of Systems with Time-Varying Delays}
\label{subsec:ODEPDE_for_timevar_delays}
Now, let us consider a system with a time-varying delay. To account for the time-varying nature of $\tau(t)$, we use an integral transformation of the delay term. Specifically, we reformulate the delay in terms of the variable $\phi$ from the solution of the ODE-PDE in~\eqref{eqn:PDE-solution} as \vspace{-2.5mm}
\begin{align*}
& x(t -\tau(t)) = x(t) - \int\nolimits_{-\tau(t)}^0 \hspace{-6mm}\dot x(t + s) ds = x(t) - \int\nolimits_{-{\tau(t)}/{\tau_{max}}}^0 \hspace{-12mm} \partial_s \phi(t, s) ds. \\[-8mm]
\end{align*} 
Next, we move the dependence of the integral limits on the time-delay $\tau(t)$ into the integrand as follows  \vspace{-1mm} 
\begin{align} \label{eqn:time-delay-using-indicator}
x(t -\tau(t)) &= x(t) - \int_{-1}^0 I\bigg(s, \frac{\tau(t)}{\tau_{max}}\bigg)  \partial_s \phi(t, s) ds,
\end{align} 
where $I(s, \tau)$ is the indicator function\vspace{-2mm}
    %
\[
I(s, \tau)  = \begin{cases}
        1, & \; \text{if } s + \tau \geq 0\\
        0, & \; \text{otherwise.}
    \end{cases} \vspace{-2mm}
\]
 
 Thus, the indicator function approach as in Eqn.~\eqref{eqn:time-delay-using-indicator} allows us to propose an ODE-PDE system with a time-varying parameter to represent the DDE in~\eqref{eqn:DDE} as \vspace{-2mm}
 \begin{align}
&\dot { x}(t)   = (A + A_d)  x(t) - A_d \int_{-1}^0  I\bigg(s, \frac{\tau(t)}{\tau_{max}}\bigg) \partial_s  \phi(t, s) ds \notag\\[-1mm]
&\qquad\qquad\qquad\qquad\qquad\qquad\qquad+ B_u u(t)  + B_w w(t) \notag \\[-3mm]
&\partial_t  \phi(t, s)  = \frac{1}{\tau_{max}} \partial_s  \phi(t, s),\quad\phi(t, 0) =  x(t) \label{eqn:ODE-PDE-TVD} \\ 
& e(t) = Cx(t),  \quad  x(0) = x_0(0),  \quad \phi(0, s) = x_0(\tau_{max} s),  \notag   \\[-7mm] \notag
\end{align}
where  $t \geq 0$ and $s \in [-1, 0]$. 

The ODE-PDE in~\eqref{eqn:ODE-PDE-TVD} and the DDE in~\eqref{eqn:DDE} have equivalent internal and input-output stability properties.\vspace{-1.5mm}

\begin{thm}\label{thm:DDE-and-ODE-PDE-are-equivalent}
Let $\tau$ be continuous, $\tau(t)  \in [0, \tau_{max}]$,   $x_0 \in C^1([-\tau_{max}, 0])$ and $u, w \in C(\R_+)$.
Then, $x \in C^1([-\tau_{max}, \infty))$ satisfies DDE~\eqref{eqn:DDE} with inputs $u, w$ and initial conditions $x_0$ if and only if $x \in C^1(\R_+)$ and $\phi \in C^1(\R_+ \times [-1, 0])$ satisfy ODE-PDE~\eqref{eqn:ODE-PDE-TVD} with inputs $u, w$ and initial conditions as in Eqn.~\eqref{eqn:ODE-PDE-TVD}.
%

\end{thm}\vspace{-2.5mm}
\begin{proof}
 1) Suppose $x \in C^1([-\tau_{max}, \infty))$ satisfies DDE~\eqref{eqn:DDE}. Define $\phi(t, s)$ as in~\eqref{eqn:PDE-solution}. Then, $\phi$ satisfies the boundary conditions for the PDE in~\eqref{eqn:ODE-PDE-TVD}. Furthermore, since $x \in C^1([-\tau_{max}, \infty))$ we have $\phi \in C^1(\R_+\times[-1, 0])$ and \vspace{-1mm}
\[
\partial_t \phi(t, s) = \partial_t x(t + \tau_{max} s) =  \frac{1}{\tau_{max}} \partial_s x(t + \tau_{max} s)  = \partial_s \phi(t, s)\vspace{-2mm}
\]
for all $t \geq 0$ and $s \in [-1, 0]$. 

Hence, $\phi$ satisfies the PDE part in Eqn.~\eqref{eqn:ODE-PDE-TVD}.

Also, by Eqn.~\eqref{eqn:time-delay-using-indicator} we have that for all $t \geq 0$ 
\vspace{-2mm} 
 \begin{align*}
\dot { x}(t)   & =Ax(t) + A_d x(t - \tau(t)) + B_u u(t) + B_w w(t) \\[-1mm]
&= (A + A_d)  x(t) - A_d \int_{-1}^0  I\bigg(s, \frac{\tau(t)}{\tau_{max}}\bigg) \partial_s  \phi(t, s) ds \\[-1mm]
&\qquad\qquad\qquad\qquad\qquad\qquad\qquad + B_u u(t) + B_w w(t). \\[-7.5mm]
\end{align*}
Thus, $x$, $\phi$ satisfy ODE-PDE~\eqref{eqn:ODE-PDE-TVD}  with inputs $u, w$ and initial conditions as in Eqn.~\eqref{eqn:ODE-PDE-TVD}.

2) Now suppose $x \in  C^1([0, \infty))$ and $\phi \in  C^1(\R\times [-1, 0])$ satisfy ODE-PDE~\eqref{eqn:ODE-PDE-TVD}. Define $\hat x$ as in Eqn.~\eqref{eqn:DDE-solution-from-ODEPDE}. Since $\phi(t, 0) = x(t)$, $\phi(0, s) = x_0(-\tau_{max} s)$ and $\phi \in  C^1(\R\times [-1, 0])$ we have that $\hat x \in C^1([-\tau_{max}, \infty))$.

Let us consider the PDE in Eqn.~\eqref{eqn:ODE-PDE-TVD}. 
Since $x \in C^1(\R_+)$, there exists a unique solution of the PDE. Thus, we have that $\phi(t, s)$ has the form as in Eqn.~\eqref{eqn:PDE-solution}.

Therefore, using the properties of the indicator function, we have that $\hat x$ satisfies the DDE in~\eqref{eqn:DDE} with inputs $u, w$ and initial conditions $x_0$.
\end{proof}\vspace{-1.5mm}

\textbf{Remark 2.} Note that the ODE-PDE representation in~\eqref{eqn:ODE-PDE-TVD} could be altered to eliminate the PDE dynamics using a model-transformation approach (using $\phi_s(t,s)=\dot x(t+s)=Ax(t+s)+A_d x(t+s-\tau(t))$) as in~\cite{kharitonov2003additional}. However, such an approach introduces additional dynamics and negates equivalence of stability properties, as shown in~\cite{kharitonov2003additional}.

In the next section, we consider the time-varying delay, $\tau(t)$, as an uncertain time-varying parameter. Then,  we reformulate the ODE-PDE in~\eqref{eqn:ODE-PDE-TVD} as a feedback interconnection  between an ODE-PDE system with constant coefficients and an uncertainty operator.  
  


  \section{A Feedback Interconnection for Systems with Uncertain Bounded Time-Varying Delays} 
\label{sec:feedback_interconnection}

In this section, we assume that the time-varying delay $\tau$ is continuous with known upper and lower bounds, $\tau \in \mbf T$,  \vspace{-2mm}
\begin{equation} \label{eqn:set_of_tau}
\mbf T =\{\tau\in C(\R) \;|  \;
             \tau(t) {\in} [\tau_{min}, \tau_{max}] \text{ for all } t \geq 0\}. \vspace{-2mm}
\end{equation}

Then, we propose two approaches to represent the ODE-PDE system in~\eqref{eqn:ODE-PDE-TVD} as a feedback interconnection. The first approach (Subsec.~\ref{subsec:feedback_representation}) isolates the indicator function from the ODE-PDE system and considers the multiplicative uncertainty with infinite-dimensional input and output channels. 
In the second approach (Subsec.~\ref{subsec:feedback_representation2}), we combine the multiplicative uncertainty with the integral. We also show that the uncertainty operators are norm-bounded. Next, having represented the ODE-PDE in Eqn.~\eqref{eqn:ODE-PDE-TVD} as a system with the uncertainty, we show that the nominal system in both approaches can be described using Partial Integral Equations. Finally, in Thm.~\ref{thm:Hinf_norm} we design $H_\infty$ control for the nominal system in PIE representation.
 \begin{figure}[t]
\centering
\vspace{2mm}

\includegraphics[width = 0.8\linewidth, trim = {0 36 0 50}, clip]{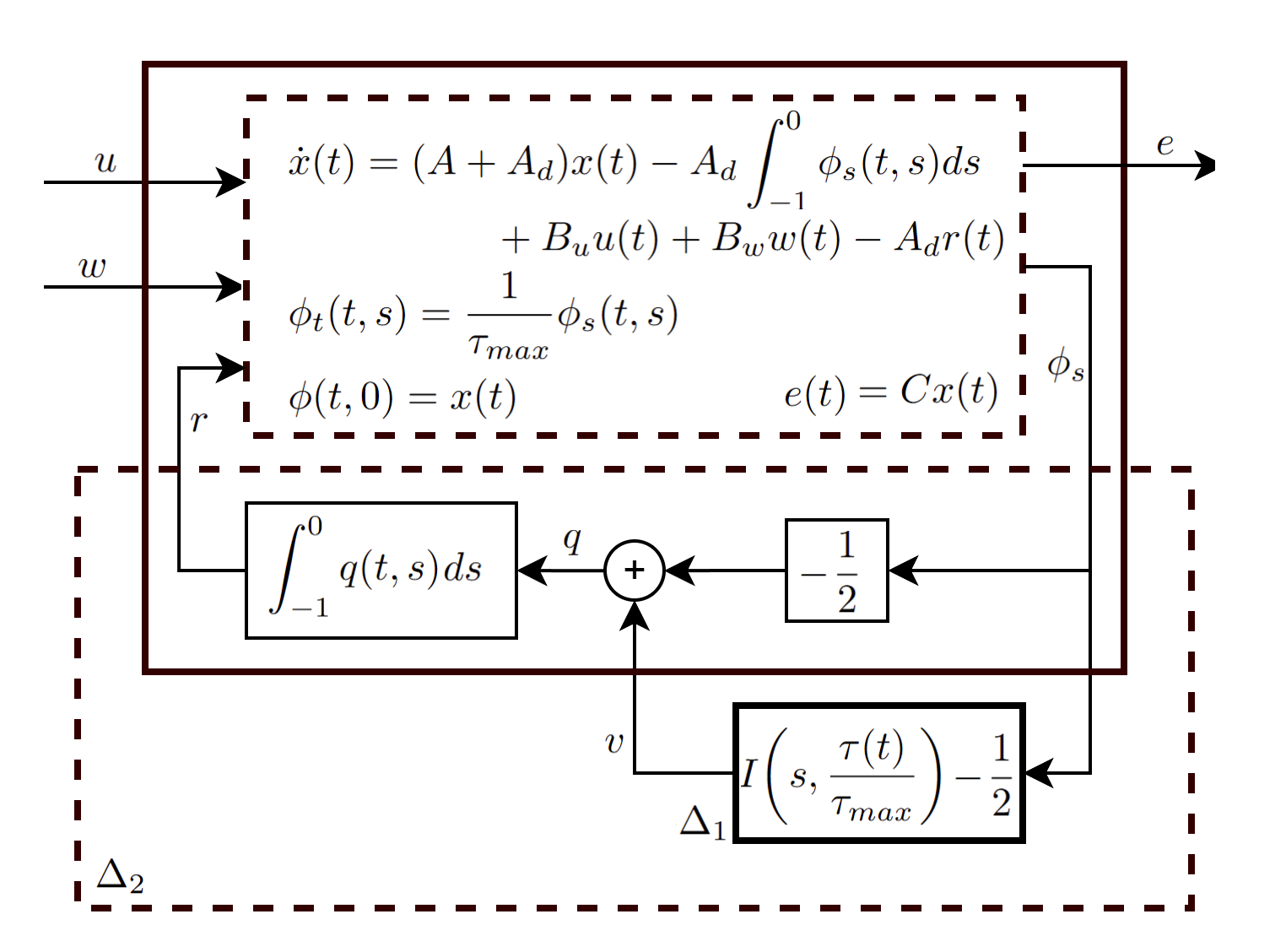}\vspace{-2mm}
\caption{The feedback interconnection representation of the ODE-PDE system in Eqn.~\eqref{eqn:ODE-PDE-TVD}. The first feedback interconnection is defined by the uncertainty operator representation $\Delta_1$  and the nominal system as in~\eqref{eqn:nominal_system} with interconnected signals $\phi_s$ and $v$. The second feedback interconnection is defined by the uncertainty operator representation $\Delta_2$ and the nominal system in~\eqref{eqn:nominal_system2}, with interconnected signals signals $\phi_s$ and $r$. The dashed blocks represent the second interconnection. 
}\vspace{-5mm}
\label{fig:interconnection}
\end{figure} 
\subsection{Feedback Interconnection with Multiplicative Uncertainty}
 \label{subsec:feedback_representation}
Let us isolate the indicator function from the ODE-PDE system in Eqn.~\eqref{eqn:ODE-PDE-TVD}. Specifically, we define the uncertainty operator $\Delta_1:L_2(\R\times[-1, 0]) \rightarrow L_2(\R\times[-1, 0])$ as\vspace{-1mm}
    \begin{equation}
      v(t,s) = (\Delta_1 z)(t, s) = \bigg[I\bigg(s, \frac{\tau(t)}{\tau_{max}} \bigg) - \frac{1}{2}\bigg] z(t,s) \label{eqn:uncertainty_operator}\vspace{-1mm}
    \end{equation} 
    and the uncertainty set is given as\vspace{-1mm}
    \begin{equation}\label{eqn:uncertainty_set}
    \mbf \Delta_1 = \bigg\{ I\bigg(s, \frac{\tau(t)}{\tau_{max}}\bigg) - \frac{1}{2} \bigg| \; \tau \in \mbf T \bigg\}.\vspace{-1mm}
    \end{equation} 
    Then, the feedback interconnection has the form as in Fig.~\ref{fig:interconnection}, where the nominal system is described by the ODE-PDE system of the form \vspace{-3mm}
        \begin{align} 
 &\dot x(t) =(A+A_d)x(t)  - \frac{1}{2}A_d \hspace{-0.5mm}\int_{-1}^0 \hspace{-2mm} \phi_s(t,s) ds - A_d \int_{-1}^0 \hspace{-2mm} v(t, s)ds  \notag \\[-1mm]
&\qquad\qquad\qquad\qquad\qquad\qquad\qquad + B_u u(t) + B_w w(t)  \notag\\[-2mm]
    &   \phi_t(t, s) =  \frac{1}{\tau_{max}}  \phi_s(t, s), \qquad s \in [-1, 0] \label{eqn:nominal_system}\\
&\phi(t, 0)  = x(t),\qquad  
        z(t, s)  = \phi_s(t, s), \qquad e(t) = C x(t),  \notag  \\[-7mm]\notag
    \end{align}
where $v(t,s) \in \R^{n_x}$ is the disturbance and $z(t, s) \in \R^{n_x}$ is the additional output.      
Then, the uncertainty operator $\Delta_1$ in Eqn.~\eqref{eqn:uncertainty_operator} is norm-bounded as shown in the next lemma. \vspace{-1.5mm}
     
    \begin{lem}\label{lem:uncertainy_is_normbounded} Let  $\tau_{max} \geq \tau_{min}  \geq 0$ and $\mbf T$ be defined as in Eqn.~\eqref{eqn:set_of_tau}. Define $\mbf \Delta_1$ as in Eqn.~\eqref{eqn:uncertainty_set}. Then, for any $\Delta_1 \in \mbf \Delta_1$, we have $\| \Delta_1 z\|_{L_2} \leq 0.5 \|z\|_{L_2} $ for all $z \in  L_2(\R\times[-1, 0])$.
\end{lem}\vspace{-4mm}
\begin{proof} 
Let $z \in L_2(\R\times[-1, 0])$ and $v(t, s) = (\Delta_1 z)(t, s)$. Since $\big|I\big(s, \frac{\tau(t)}{\tau_{max}}\big)  - \frac{1}{2}\big|< \frac{1}{2}$, we have $|v(t, s)|\leq \frac{1}{2}|z(t, s)|$ for all $t \in \R$ and $s \in [-1, 0]$. Thus, we have $\|\Delta_1 z\|_{L_2} \leq \frac{1}{2}\|z\|_{L_2}$ for all $z \in  L_2(\R\times[-1, 0])$.
\end{proof} \vspace{-1.5mm}
 
Now, let us consider an alternative feedback interconnection for the ODE-PDE system. \vspace{-1.5mm}


 \subsection{Feedback Representation with Integral Uncertainty}
 \label{subsec:feedback_representation2}
In Lem.~\ref{lem:uncertainy_is_normbounded}, we exploited a bound on $\norm{I(\bullet,\frac{\tau(t)}{\tau_{max}})-\frac{1}{2}}_{L_\infty} $ to obtain an induced norm bound on the uncertain Multiplication operator, $\Delta_1$. 
 In this subsection, we utilize a bound on $\norm{I(\bullet,\frac{\tau(t)}{\tau_{max}}) - 1}_{L_2}$. However, bounds on the $L_2$-norm of a multiplier do not translate directly to bounds on the induced norm. Hence, we shift a spatial integral into the uncertainty and interpret this as the inner product of $I(s,\frac{\tau(t)}{\tau_{max}}) - 1$ with the input, $z$. 
 
Specifically, let us define a modified uncertainty operator, $\Delta_2: L_2(\R\hspace{-0.5mm}\times\hspace{-0.5mm} [-1, 0]) \rightarrow L_2(\R)$, of the form \vspace{-1mm}
\begin{equation}
v(t) = (  \Delta_2 z)(t) := \int_{-1}^0 \bigg[I\bigg(s,  \frac{\tau(t)}{\tau_{max}}\bigg) - 1\bigg] z(t, s) ds.\label{eqn:uncertainty_operator2}\vspace{-1mm}
\end{equation}


For $\tau \in \mbf T$, we define a set of uncertainty operators as\vspace{-1mm}
%
%
\begin{equation}\label{eqn:uncertainty_set2}
 {\mbf \Delta}_2 := \big\{ \Delta_2  \text{ as in Eqn.~\eqref{eqn:uncertainty_operator2}} \;\big|\;\tau \in \mbf T \big\}.\vspace{-1mm}
\end{equation} 

Then, the nominal system has the form  \vspace{-1.5mm}
        \begin{align} 
 &\dot x(t)  \hspace{-0.5mm}= \hspace{-0.5mm}(A \hspace{-0.5mm}+ \hspace{-0.5mm}A_d)x(t)   - A_d \hspace{-0.5mm} \int_{-1}^0 \hspace{-2mm} \phi_s(t, s) ds  \hspace{-0.5mm}-  \hspace{-0.5mm}A_d v(t) \notag\\[-2mm]
 	& \qquad\qquad\qquad\qquad\qquad\qquad   +  B_u u(t)  +  B_w w(t) \notag\\[-2mm]
      & \phi_t(t, s) =  \frac{1}{\tau_{max}}  \phi_s(t, s), \qquad s \in [-1, 0] \label{eqn:nominal_system2}\\
       &\phi(t, 0)  = x(t),\qquad  
        z(t, s)  = \phi_s(t, s), \qquad
        e(t) = C x(t),  \notag \\[-7mm]\notag   \end{align}
    where  $v(t) \in \R^{n_x}$ is the disturbance and $z(t, s) \in \R^{n_x}$ is the output of the system. 
    
Although $\Delta_2$ is defined as an inner-product (projection) as opposed to multiplicative uncertainty, we may still use the $L_2$ norm bound on parameter $\norm{I(\bullet,\frac{\tau(t)}{\tau_{max}}) - 1}_{L_2} \leq  \sqrt{\frac{\tau_{max}-\tau_{min}}{\tau_{max}}}$ to bound the operator norm of $\Delta_2$ as follows.
%

\begin{lem}\label{lem:uncertainy_is_normbounded2} Let  $\tau_{max} \geq \tau_{min}  \geq 0$ and $\mbf T$ be defined as in Eqn.~\eqref{eqn:set_of_tau}.  Define $\mu = \sqrt{\frac{\tau_{max} - \tau_{min}}{\tau_{max}}}$ and $\mbf \Delta_2$ as in Eqn.~\eqref{eqn:uncertainty_set2}.

Then, for any $ \Delta_2 \in   {\mbf \Delta_2}$, we have $\|  \Delta_2 z\|_{L_2} \leq \mu\|z\|_{L_2} $ for any $z \in  L_2(\R\times[-1, 0])$.
\end{lem}\vspace{-2mm}
\begin{proof}
  
For any $ \Delta_2 \in \mbf \Delta_2$ we have \vspace{-4mm}
{\small
 \begin{align*}
\|(\Delta_2 z)(t)\|^2_2 &\hspace{-0.5mm}=\hspace{-0.5mm}  \bigg\|  \int\limits_{0}^{-1} \hspace{-1mm}\bigg[I\bigg(\hspace{-0.5mm}s, \frac{\tau(t)}{\tau_{max}}\hspace{-0.5mm}\bigg)\hspace{-1mm} -\hspace{-1mm} 1\bigg]z(t, s)ds  \bigg\|_2^2 \hspace{-0.5mm}=\hspace{-0.5mm}\bigg\| \hspace{-3mm} \int\limits_{-1}^{-\frac{\tau(t)}{\tau_{max}}} \hspace{-3mm} z(t, s)ds  \bigg\|_2^2 
 \end{align*}} \vspace{-4mm}
 
 for all $z \in L_2(\R\times[-1, 0])$ and $t \geq 0$.

Next, by Jensen inequality (\cite{gu2003stability}, p. 322) we have \vspace{-2mm}
{
\begin{align*}
 \|( \Delta_2 z)(t)\|_2^2 &\leq \left(1-\frac{\tau(t)}{\tau_{max}}\right)   \int_{-1}^{-\frac{\tau(t)}{\tau_{max}}}  \|z(t, s) \|_2^2 ds\\[-1mm]
 &\hspace{-15mm}\leq \left(1-\frac{\tau_{min}}{\tau_{max}}\right) \int_{-1}^{-1 + \mu^2} \hspace{-3mm} \|z(t, s)\|^2_2 ds \leq \mu^2 \int_{-1}^0 \|z(t, s)\|^2_2 ds \\[-7mm]
\end{align*}}
 for any $z \in L_2(\R\times[-1, 0])$, $t \geq 0$ and $\tau \in \mbf T$. 

Therefore, for any $\Delta_2 \in  {\mbf \Delta_2}$ we have   $\|\Delta_2 z\|_{L_2} \leq \mu \|z\|_{L_2}$ for any $z \in L_2(\R\times [-1, 0])$. \vspace{-2mm}
\end{proof}

\textbf{Remark 3.} To interpret the uncertainty operator $\Delta_2$, we note that when $z(t,s)=\phi_s(t,s)=\dot x(t + \tau_{max} s)$,  Eqn.~\eqref{eqn:uncertainty_operator2} reduces to $(\Delta_2 \dot x)(t) = x(t - \tau(t)) - x(t - \tau_{max})$. However, the uncertainty operator in~\eqref{eqn:uncertainty_operator2} is slightly more general in that it does not presume $z(t, s) = \dot x(t + \tau_{max} s)$.\vspace{-1.5mm}

\subsection{PIE Representation for the Nominal System}
\label{subsec:nominal}

Now,  let us describe the nominal system as in Eqn.~\eqref{eqn:nominal_system} or~\eqref{eqn:nominal_system2} using Partial Integral Equations. For example as discussed in~\cite{peet2021representation}, for given matrix $P$, bounded functions $Q_1, Q_2, R_0$ and separable functions $R_1(s, \theta), R_2(s, \theta)$ we say $\mcl P$ is a Partial Integral (PI) operator on $\mbf Z^{n,m} := \R^{n} \times L_2^{m}[a, b]$ if 

\vspace{-4mm}{\small
 \[
 \bigg(\mcl P\bmat{P & Q_1 \\ Q_2 & \{R_i\}} \bmat{x \\ \mbf \Phi}\bigg)(s) = \bmat{Px + \int_{a}^b Q_1(s) \mbf \Phi(s) ds \\ 
 Q_2(s) x + (\mcl P_{R_i} \mbf \Phi)(s)},
 \]}
 where $(\mcl P_{R_i} \mbf \Phi)(s) = R_0(s) \mbf \Phi(s) + \int_{a}^s R_1(s, \theta) \mbf \Phi(\theta) d\theta + \int_{s}^b R_2(s, \theta) \mbf \Phi(\theta)$.
Then, the ODE-PDE systems as in Eqn.~\eqref{eqn:nominal_system}  or Eqn.~\eqref{eqn:nominal_system2} can be represented as  \vspace{-2mm}
   \begin{align} 
&\partial_t(\mcl T \mbf x(t))  = \mcl A \mbf x(t) + \mcl B_{v} \bmat{w(t) \\ \mbf v(t)} + \mcl B_u u(t)  \notag\\[-3mm]
& \bmat{e(t) \\ \mbf z(t)}  = (\mcl C \mbf x)(t),  \label{eqn:PIE} \\[-7mm]\notag
\end{align} 
 where $t \geq 0$, $\mbf x(t) \in \mbf Z^{n_x, n_x}$ is the states  and $\mcl T, \mcl A, \mcl B_u, \mcl B_v, \mcl C$ are PI operators with appropriate dimensions.
 
 Specifically, to represent the nominal system in Eqn.~\eqref{eqn:nominal_system} the PI operators can be chosen as \vspace{-4mm}

{\small
 \begin{align}
& \mcl C  = \mcl P \bmat{C & 0 \\ 0 & \{I, 0, 0\}} \quad\hspace{-2.5mm} \mcl B_u = \mcl P\bmat{B_u &\emptyset \\ 0  &\emptyset} \quad\hspace{-2.5mm}  \mcl B_v =  \mcl P\bmat{B_w & - A_d \\ 0 & 0} \notag\\
& \mcl T  = \mcl P\bmat{I & 0 \\ I & \{0, 0, -I\}} \hspace{1mm} \mcl A = \mcl P\bmat{A\hspace{-0.75mm} + \hspace{-0.75mm}A_d & -\frac{1}{2}A_d \\ 0 & \{\frac{1}{\tau_{max}} I, 0, 0\}}.\notag 
 \\[-6mm]\label{eqn:PI_operators}
\end{align}  } 
For the second representation in Subsec.~\ref{subsec:feedback_representation2}, the nominal system in Eqn.~\eqref{eqn:nominal_system2} can be represented as PIE~\eqref{eqn:PIE} using the PI operators of the form \vspace{-3mm}

{\small
 \begin{equation}
\mcl A = \mcl P\bmat{A\hspace{-0.75mm} + \hspace{-0.75mm}A_d & -A_d \\ 0 & \{\frac{1}{\tau_{max}} I, 0, 0\}}, \; \mcl B_v =  \mcl P\bmat{[B_w \; -A_d] &  \emptyset \\ 0 &  \emptyset} \label{eqn:PI_operators2} \vspace{-1mm}
\end{equation} }
and $\mcl T, \mcl B_u$ and $\mcl C$ as in Eqn.~\eqref{eqn:PI_operators}.

Then, the states of the PIE in Eqn.~\eqref{eqn:PIE} represent the states of the nominal systems in~\eqref{eqn:nominal_system} or~\eqref{eqn:nominal_system2} as  \vspace{-2mm}
 \begin{equation} 
\mbf x(t, s) = \bmat{x(t) & \partial_s \phi(t,s)}^T. \label{eqn:PI_variables_to_PDE} \vspace{-2mm}
\end{equation}

 Then, input-output maps represented by the ODE-PDE in~\eqref{eqn:nominal_system} and the PIE in~\eqref{eqn:PIE} are equivalent as established in~\cite{peet2021representation}. Furthermore, the $H_\infty$ state-feedback controller for the PIE in~\eqref{eqn:PIE} can be integrated in the PIE framework.\vspace{-1mm}

\begin{thm}[Thm. 11 and Cor.~14 in~\cite{shivakumar2022dual}]~\label{thm:Hinf_norm}
Let PI operators $\mcl T, \mcl A, \mcl B_u, \mcl B_v, \mcl C_z$ be PI operators. Suppose there exist $\mu > 0$, PI operators $\mcl Q, \mcl R:\mbf Z^{n_x, n_x} \hspace{-0.5mm}\rightarrow \hspace{-0.5mm}\mbf Z^{n_x, n_x}$, $\mcl Z:\mbf Z^{n_x, n_x}\hspace{-0.5mm} \rightarrow\hspace{-0.5mm} \mbf Z^{n_u, 0}$, such that $\mcl T \mcl Q^* = \mcl Q \mcl T^* = \mcl R \geq 0$, $\mcl Q$ is invertible, and 

{\small \vspace{-4mm}
\begin{equation}\label{eqn:LPI_Hinf_controller}
\bmat{-\frac{1}{\mu} \mcl I & 0 & \mcl C \mcl Q^* \\ 
		0 & -\frac{1}{\mu} \mcl I & \mcl B_v \\ 
		 \mcl Q  \mcl C^*& \mcl B_v^* & (\mcl A \mcl Q + \mcl B_u\mcl Z) + (\mcl A \mcl Q + \mcl B_u\mcl Z)^*} \preceq 0.
\end{equation}}\vspace{-2mm}

Define $\mcl K = \mcl Z \mcl Q^{-1}$ and $u(t) = \mcl K \mbf x(t)$ for all $t\geq 0$. 
Then,  we have that $
(\|e\|_{L_2} + \|\mbf z\|_{L_2})  \leq \frac{1}{\mu} (\|w\|_{L_2} + \|\mbf v\|_{L_2} ) $ 
for any $\mbf v,\mbf z \hspace{-0.5mm}\in\hspace{-0.5mm} L_2(\R \times [-1, 0])$ and $e, w\hspace{-0.5mm} \in\hspace{-0.5mm} L_2(\R)$ satisfying the PIE in~\eqref{eqn:PIE} with $\mbf x(0) =0$. 
\end{thm} \vspace{-2mm}
Having defined the feedback interconnection between the nominal system described using PIEs and the uncertainty, the task is to derive sufficient conditions for the robust stabilization of the interconnected system. 
 
\section{Stabilization of Systems with Time-Varying Delays using PIE}\label{sec:stabilization}
 
 In Sec.~\ref{sec:feedback_interconnection}, we have shown that the proposed ODE-PDE representations in Eqn.~\eqref{eqn:ODE-PDE-TVD} can be represented as a feedback interconnection between a norm-bounded uncertainty operator as in Eqn.~\eqref{eqn:uncertainty_operator} or~\eqref{eqn:uncertainty_operator2} and a nominal system represented in the PIE format as in Eqn.~\eqref{eqn:PIE}. The goal of this section is to consider the robust stabilization problem for the PIE system with uncertainty. Specifically, we consider the system \vspace{-2mm}
\begin{align}
\partial_t (\mcl T \mbf x(t)) &= \mcl A x(t) + \mcl B_v  \bmat{w(t) \\ \mbf v(t)}  + \mcl B_u u(t), \quad u(t)   = \mcl K \mbf x(t)  \notag \\[-2mm] 
\bmat{e(t) \\ \mbf z(t)}  &= \mcl C \mbf x(t), \quad
\mbf v(t) = (\Delta \mbf z)(t), \quad \Delta \in \mbf \Delta, \label{eqn:PIE_with_uncertainty}  \\[-7mm] \notag
\end{align}
 where $\mbf x(t) \in \mbf Z^{n_x, m_x}$ is the states, $e(t) \in \R^{n_e}$ is the output, $w(t) \in \R^{n_w}$ is the exogenous input, $\mbf z(t) \in \mbf Z^{n_z, m_z}$ is the disturbance output, $\mbf v(t) \in \mbf Z^{n_v, m_v}$ is the disturbance, $u(t) \in \R^{n_u}$ is the control input. $\mcl K$ is the control gain and $\mcl T, \mcl A, \mcl B_v, \mcl C, \mcl B_u$ are PI operators with appropriated dimensions. The uncertainty set $\mbf \Delta$ is a set of norm-bounded operators of the form  \vspace{-3mm}
  \begin{align} \label{eqn:uncertainty_set_PI}
\hspace{-2.5mm} \mbf \Delta =\{\Delta\hspace{-0.5mm}:\hspace{-0.5mm}\mbf Z^{n_z, m_z} \rightarrow \mbf Z^{n_v, m_v} \; |\;  \|\Delta\|\leq \mu\}. \\[-8mm] \notag
 \end{align}  
 
In the next theorem, we provide sufficient conditions for $L_2$ stability of the PIE system in Eqn.~\eqref{eqn:PIE_with_uncertainty}.\vspace{-2mm}

\begin{thm}[Stabilizing Controller]\label{thm:PIE_control}
Let $\mu \geq 0$ and the set $\mbf \Delta$ be given as in Eqn.~\eqref{eqn:uncertainty_set_PI}. Suppose there exist $\delta < \frac{1}{\mu}$ and PI operators $\mcl Q, \mcl R:\mbf Z^{n_x, n_x} \rightarrow \mbf Z^{n_x, n_x}$, $\mcl Z:\mbf Z^{n_x, n_x} \rightarrow \mbf Z^{n_u, 0}$, such that $\mcl T \mcl Q^* = \mcl Q \mcl T^* = \mcl R \geq 0$, $\mcl Q$ is invertible, and \vspace{-4mm}

{\small
\begin{equation}\label{eqn:LPI_robust_controller}
\hspace{-1mm}\bmat{-\delta \mcl I & 0 & \mcl C \mcl Q^* \\ 
		0 & -\delta \mcl I & \mcl B_v \\ 
		 \mcl Q \mcl C^*& \mcl B_v^* &
		 (\mcl A \mcl Q + \mcl B_u\mcl Z)  + (\mcl A \mcl Q + \mcl B_u\mcl Z)^*  
		 } \hspace{-1mm} \preceq  \hspace{-0.5mm}0.
\end{equation}\vspace{-2mm}}

Define $\mcl K = \mcl Z \mcl Q^{-1}$. Then,  we have that
$\|e\|_{L_2} \leq \frac{\delta}{1 - \mu\delta} \|w\|_{L_2}$ for any $e, w \hspace{-0.5mm}\in\hspace{-0.5mm} L_2(\R)$ satisfying the PIE in~\eqref{eqn:PIE_with_uncertainty} with $\mbf x(0)\hspace{-0.5mm} = \hspace{-0.5mm}0$.
\end{thm}\vspace{-4mm}
\begin{proof}
By Thm.~\ref{thm:Hinf_norm} we have that $(\|e\|_{L_2} + \|\mbf z\|_{L_2})  \leq \delta (\|w\|_{L_2} + \|\mbf v\|_{L_2} )$ 
for any $\mbf v,\mbf z \hspace{-0.5mm}\in\hspace{-0.5mm} L_2(\R \times [-1, 0])$ and $e, w\hspace{-0.5mm} \in\hspace{-0.5mm} L_2(\R)$ satisfying the PIE in~\eqref{eqn:PIE} with $\mbf x(0) =0$. Furthermore, for any $\Delta \in \mbf \Delta$ we have $\|\mbf v\|_{L_2} \leq \mu \|\mbf z\|_{L_2}$.

 Thus, since $\delta \mu < 1$, the Small-Gain Theorem~\cite{gu2003stability} implies that $\|e\|_{L_2}  \leq \frac{\delta}{1 - \delta\mu}  \|w\|_{L_2}$ 
for any $w, e \in L_2(\R)$ satisfying the PIE in~\eqref{eqn:PIE_with_uncertainty} with $\mbf x(0) =0$.
\end{proof}\vspace{-3mm}
 
Using the PIE formulation of the nominal system in Eqn.~\eqref{eqn:nominal_system} or~\eqref{eqn:nominal_system2} and describing the uncertainty as in Eqn.~\eqref{eqn:uncertainty_operator} or~\eqref{eqn:uncertainty_operator2}, we may use the Thm.~\ref{thm:PIE_control} to design a stabilizing controller for the DDE in~\eqref{eqn:DDE_control}. \vspace{-2mm}

\begin{cor}\label{cor:stabilization1}
Let $\tau_{max} > 0$, $\tau_{min} \in [0, \tau_{max}]$ and $\mbf T$ be defined as in Eqn.~\eqref{eqn:set_of_tau}. Suppose that either of the following statements hold. \vspace{-1.5mm}
\begin{enumerate}
\item [a)] Conditions of Thm.~\ref{thm:PIE_control} are satisfied for PI operators as in Eqn.~\eqref{eqn:PI_operators} and $\mu = \frac{1}{2}$.\vspace{-0.5mm}
\item [b)] Conditions of Thm.~\ref{thm:PIE_control} are satisfied for PI operators as in Eqn.~\eqref{eqn:PI_operators2} and  $\mu = \sqrt{\frac{\tau_{max} - \tau_{min}}{\tau_{max}}}$.\vspace{-1mm}
\end{enumerate}
Then, if $e, w\hspace{-0.5mm} \in\hspace{-0.5mm} L_2(\R)$ satisfy the DDE in Eqn.~\eqref{eqn:DDE} with control $u(t) = \mcl K\bmat{x(t) \\\partial_s  x(t + \tau_{max} s) }$ and $x_0 = 0$, we have $\|e\|_{L_2}\hspace{-0.5mm} \leq\hspace{-0.5mm}  \frac{\delta}{1 - \mu\delta}\|w\|_{L_2}$. 
\end{cor} 
 \vspace{-2mm}
\begin{proof} \vspace{-2mm}

Suppose that $a)$ holds. Then,  as shown in
Thm.~\ref{thm:DDE-and-ODE-PDE-are-equivalent}, if $e, u, w$ and $x$ satisfy the DDE in Eqn.~\eqref{eqn:DDE}, then the $u, w$, $x$  satisfy the ODE-PDE system in Eqn.~\eqref{eqn:ODE-PDE-TVD} with $\phi$ as in Eqn.~\eqref{eqn:PDE-solution}. Furthermore, the nominal ODE-PDE system in~\eqref{eqn:nominal_system} and the PIE system in Eqn.~\eqref{eqn:PIE} with the PI operators as in~\eqref{eqn:PI_operators} have equivalent input-output properties as shown in~\cite{peet2021representation}. Furthermore, if $x$ is the solution of the DDE, then for the PIE in~\eqref{eqn:PIE_with_uncertainty} with the uncertainty as in~\eqref{eqn:uncertainty_operator} the solution has the form $\mbf x(t, s) = \bmat{x(t) & \partial_s x(t + \tau_{max} s)}^T$.


Thus, if $e, w$ and $x$ satisfy the DDE in~\eqref{eqn:DDE} with control $u(t) = \mcl K\bmat{x(t) \\ \partial_s x(t + \tau_{max} s) }$, then $e, w$ and $\mbf x$ satisfy the PIE in~\eqref{eqn:PIE_with_uncertainty}.  Finally, since the uncertainty operator in~\eqref{eqn:uncertainty_operator} is norm-bounded by $\mu$ (Lem.~\ref{lem:uncertainy_is_normbounded}) and conditions of Thm.~\ref{thm:PIE_control} are satisfied, we have $\|e\|_{L_2} \hspace{-1mm} \leq \hspace{-1mm} \frac{\delta}{1 - \mu\delta}\|w\|_{L_2}$. The proof for $b)$ is similar. \vspace{-4mm}

\end{proof}\vspace{-2mm}

Cor.~\ref{cor:stabilization1} provides sufficient conditions for stabilization of the DDE with time-varying delays. In the next section, we apply Cor.~\ref{cor:stabilization1} a) to systems with $\tau_{min} = 0$ and Cor.~\ref{cor:stabilization1} b) to systems with $\tau_{min} > 0$.
\
\section{Numerical Examples}\label{sec:examples}

In the following examples, we consider the problem of stabilization of time-delay systems with uncertain time-varying delays. For all examples, we first transform the system to the PIE reformulation as defined in Eqn.~\eqref{eqn:PIE}. Then, using PIETOOLS~\cite{shivakumar2022dual}, we use Cor.~\ref{cor:stabilization1} to design a stabilizing controller for the original system. \vspace{-2mm}

\subsection{Example 1}\vspace{-1mm}
In the first example, we consider a modified example from~\cite{gu2003stability}. Specifically, we consider the DDE in~\eqref{eqn:DDE} where\vspace{-2mm}
\begin{align*}\small
A = \bmat{-2 & 0 \\ 0 & -0.9}, \quad A_d = \bmat{-1 & 0 \\ -1 & -1}, \quad B_u = \bmat{1 \\ 1}, \\[-0.5mm]
B_w = B_u, \quad C = \bmat{0& 1}. \\[-7mm]
\end{align*}  
The nominal system is stable for fixed delay $\tau \in [0,6.17]$. The task is to find the largest interval $[\tau_{min}, \tau_{max}]$ and control law $K_1, K_2$ such that the DDE in~\eqref{eqn:DDE} with control law as in~\eqref{eqn:DDE_control} is stable for known $\tau_{min}$. 

\textit{a)} If $\tau_{min} = 0$, then using Cor.~\ref{cor:stabilization1} a), we were able to find $K_1, K_2(s)$ such that the closed-loop system has a finite $L_2$-gain  for all $\tau(t) \in [0, 2.6]$, which is an improvement over an implementation of the conditions in~\cite{briat2010memory} (stable for $\tau(t) \in [0, 1]$). The synthesized controller gains are $K_1 = [-413.2001,-71.9607]$ and\vspace{-2.5mm} 
\[\small
  K_2(s) = \bmat{48.38 s^4 + 1273.47 s^3 - 8.2s^2 
  - 1846.9s - 760.4\\ 
   37.8s^4 + 1204.1s^3 + 66.7s^2 
  - 1832.3s - 990.4}^T.\vspace{-2mm}
  \]\vspace{-2.5mm}

\textit{b)} For $\tau_{min} = 6.2$ we find controller gains $K_1, K_2(s)$ using Cor.~\ref{cor:stabilization1} b), such that the closed-loop system has a finite $L_2$-gain for $\tau(t) \in [6.2, 6.48]$ -- an improvement over an implementation of the conditions from~\cite{briat2010memory} ($\tau(t) \in [6.2, 6.35]$).  The resulting controller gains are $K_1 = [-287.18,272.55]$ and \vspace{-2.5mm}
\[\small
  K_2(s) = \bmat{-6.5\cdot 10^{-8}s^2 - 190.70s - 141.9 \\ 
    - 381.02s - 462.07}^T.\vspace{-2mm}
  \] \vspace{-2.5mm}

\subsection{Example 2} \vspace{-1mm}
\begin{figure}[t]
\centering
\vspace{2mm}

\includegraphics[width = 0.9\linewidth, trim = {30 10 30 30}, clip]{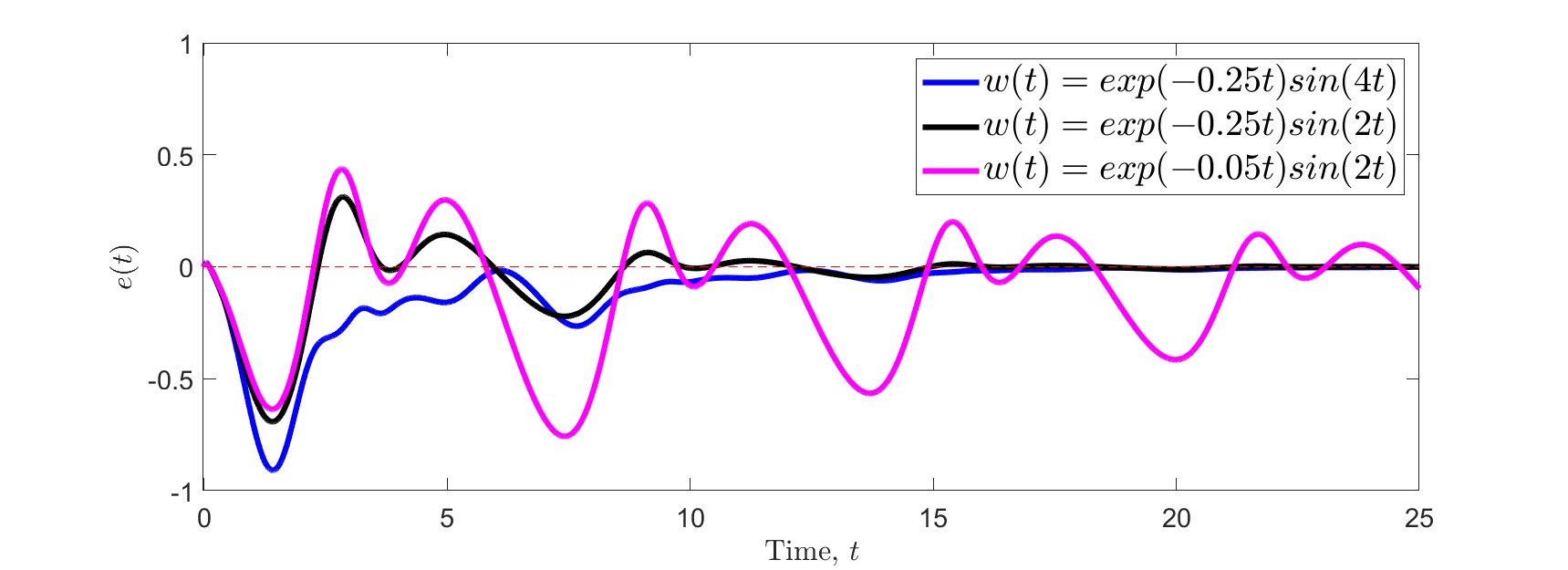}\vspace{-3mm}

\caption{ Numerical Simulation of the output of the system in Example 2a with controller gain computing using Cor.~\ref{cor:stabilization1}, exogenous inputs $w(t)$ and initial conditions $x_0 = 0$. The time-varying delay is given by $\tau(t) = 1.15(0.5 + 0.5 \sin(t)) \in [0, 1.15]$. The simulation uses trapezoidal integration with a time-step of $0.01$.}\vspace{-4mm}
\label{fig:simulation}
\end{figure} 
The second example is from~\cite{fridman2006new} (Example 3), where the DDE has the form as in Eqn.~\eqref{eqn:DDE} with  \vspace{-2mm}
\begin{align*}\small
A = \bmat{0 & 0 \\ 0 & 1}, \quad A_d = \bmat{-1 & -1 \\ 0 & -0.9}, \quad B_u = \bmat{0 \\ 1}. \\[-7.5mm]
\end{align*} 
The nominal system is stabilizable, but not stable for fixed delay $\tau \in [0,3.2)$~\cite{fridman2006new}. Our goal is  to find the largest  $\tau_{max}$ and control $K_1$, $K_2(s)$ such that DDE~\eqref{eqn:DDE} with $u$ as defined in Eqn.~\eqref{eqn:DDE_control} is stable for all $\tau(t) \in [\tau_{min}, \tau_{max}]$ for given $\tau_{min}$. 

\textit{a)} We use $\tau_{min} =  0$, $B_w = [-2\; -0.9]^T$ and $C = \bmat{0& 1}$. Using Cor.~\ref{cor:stabilization1} a), we find the closed-loop system has a finite $L_2$-gain for all $\tau(t) \in [0, 1.16]$ (improving over the bound $\tau(t) \in [0, 1)$ in~\cite{fridman2006new}). The controller gains are $K_1 = [-6.919,  -11.317]$ and \vspace{-2.5mm}
 \[\small
 \quad K_2(s) = \bmat{- 28.16 s^3 - 44.91 s^2 - 16.42 \\  - 118.94 s^3 - 210.85 s^2 - 114.08 s - 20.48 }^T.\vspace{-2.5mm}
  \] 
  A numerical simulation of the closed-loop system with the proposed control gains is given in Fig.~\ref{fig:simulation}.

\textit{b)} For $\tau_{min} = 1.8$,  $B_w = A_d$ and $C = \bmat{1 & 1}$, we use Cor.~\ref{cor:stabilization1} b) to show that the closed-loop system has a finite $L_2$-gain for all $\tau(t) \in [1.8, 2.295]$ (improving the previous results $\tau(t) \in [1.8, 2.2]$~\cite{fridman2006new}). $K_1 = [-311.5,-312.5]$ and \vspace{-2.5mm}

 \[\small
  K_2(s) = \bmat{- 151.96s^2 - 170.61s - 13.57 \\ -264.37 s^2 - 351.05 s - 85.88}^T.
  \] \vspace{-4mm}

\section{Conclusion}
In this paper, we addressed the problem of state-feedback stabilization for linear DDEs with a time-varying delay. We reformulated the DDE as a feedback interconnection between a nominal PIE system and a norm-bounded uncertainty. To construct the representation, we first considered ODE-PDE and PIE state-space representations of the DDE with time-varying delay. Such formulations allowed us to consider the DDE as a feedback interconnection of the nominal time-invariant PIE system with an infinite-dimensional time-varying parametric uncertainty. Then, using $H_\infty$ optimal control for the nominal PIE system and the small gain conditions, we proposed sufficient conditions for stabilizing the DDE with time-varying delay. Finally, the effectiveness of the method was verified using simple numerical examples.



\raggedbottom
\bibliographystyle{IEEEtran}

\bibliography{TimeDelay}

\end{document}